\documentclass[prl,prl,twocolumn,showpacs,superscriptaddress,groupedaddress]{revtex4}  
\usepackage{graphicx}  
\usepackage{dcolumn}   
\usepackage{bm}        
\usepackage{amssymb}   
\usepackage[utf8x]{inputenc}
\usepackage{amssymb,amsmath}
\usepackage{gensymb}
\usepackage{textcomp}
    
\graphicspath{{figures/}}

\begin{document}

\widetext


\title{An Underlying Asymmetry within Particle-size Segregation}


\author{K. van der Vaart}\email{kasper.vandervaart@epfl.ch}
\affiliation{Environmental Hydraulics Laboratory, \'{E}cole Polytechnique F\'{e}d\'{e}rale de Lausanne, \'{E}cublens, Lausanne, Switzerland}

\author{P. Gajjar}%
\affiliation{School of Mathematics and Manchester Centre for Nonlinear Dynamics, University of Manchester, Manchester, UK
}

\author{G. Epely-Chauvin}%
\affiliation{Environmental Hydraulics Laboratory, \'{E}cole Polytechnique F\'{e}d\'{e}rale de Lausanne, \'{E}cublens, Lausanne, Switzerland}

\author{N. Andreini}%
\affiliation{Environmental Hydraulics Laboratory, \'{E}cole Polytechnique F\'{e}d\'{e}rale de Lausanne, \'{E}cublens, Lausanne, Switzerland}

\author{J.M.N.T. Gray}%
\affiliation{School of Mathematics and Manchester Centre for Nonlinear Dynamics, University of Manchester, Manchester, UK
}%
\author{C. Ancey}%
\affiliation{Environmental Hydraulics Laboratory, \'{E}cole Polytechnique F\'{e}d\'{e}rale de Lausanne, \'{E}cublens, Lausanne, Switzerland}

\date{\today}

\begin{abstract}

We experimentally study particle scale dynamics during segregation of a bidisperse mixture under oscillatory shear. Large and small particles show an underlying asymmetry that is dependent on the local particle concentration, with small particles segregating faster in regions of many large particles and large particles segregating slower in regions of many small particles. We quantify the asymmetry on bulk and particle scales, and capture it theoretically. This gives new physical insight into segregation and reveals a similarity with sedimentation, traffic flow and particle diffusion.   


\end{abstract}
\pacs{45.70.Mg, 05.45.-a, 47.57.Gc}

\maketitle

The natural tendency of granular media to self-organize when agitated or sheared produces a rich diversity of complex and beautiful patterns~\cite{AransonTsimring2008, ottino_khakhar2000, PihlerPuzovicMullin2013}. Although it is counter-intuitive that the components of a heterogeneous mixture will readily separate, this property has serious technical implications as the cause of product non-uniformity in many industrial processes~\cite{Williams1968, johanson1978, shinbrot_muzzio2000} and also plays a pivotal role in the enhanced run-out of large scale geophysical granular flows, such as debris-flows, pyroclastic flows and snow avalanches~\cite{Iverson11007, johnson2012, PalladinoValentine11005,McElwaineNishimura2000}. A firm knowledge of the segregation process is thus of universal importance.

Although there has been considerable recent progress in developing continuum based segregation models for sheared granular flows \cite{BridgwaterFooStephens1985, gray_thornton2005, may2010, fan2014, marks2012, Hill_Tan2014}, the individual particle dynamics are still poorly understood. Discrete Particle Method (DPM) simulations \cite{bridgwater2010, thornton2012modeling, FanEtAl2013, staron2014} produce a wealth of micro-scale information, but are models in themselves. It is vital to directly measure particle segregation dynamics in real experiments, but such an analysis is difficult with conventional techniques such as binning and side-wall observation~\cite{savage_lun1988, vallance_savage2000, wiederseiner2011, golick_daniels2009}. Non-intrusive imaging techniques, such as X-ray tomography~\cite{McDonaldEtAl2012} and refractive index-matched scanning (RIMS)~\cite{dijksman2012, wiederseiner2011b} allow examination of the interior of a granular medium, with RIMS recently developing into a powerful tool for examining monodisperse and bidisperse flows~\cite{herrera2011, Slotterback2012,harrington2013}. In particular, the work of \citet{harrington2013} on the emergence of granular segregation demonstrates how particle scale analysis can give new physical insights.  

In this Letter, we analyze particle scale dynamics during segregation of a bidisperse mixture under oscillatory shear. We find that the behavior of small and large particles exhibits an asymmetry related to the local particle concentration, with small grains moving faster through regions of many large particles and large particles rising slower through regions of many small particles. This asymmetry is quantified on both particle and bulk length scales, and it is shown how to incorporate the behavior within the theoretical framework.

\begin{figure}[!b]
\resizebox{0.48\textwidth}{!}{\includegraphics{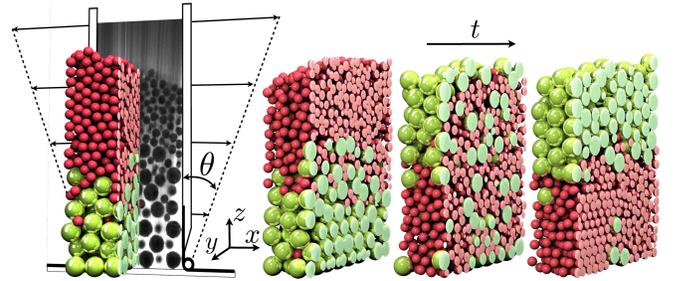}}
\caption{(color online). Left: The experimental setup. A raw data image is shown and a cross-section of a reconstructed sample with 3~mm and 6~mm beads. Right: cross-sections at different times during an experiment.}
\label{fig:setup}
\end{figure}

\emph{Methods.---}A  shearbox 51~mm deep and 37~mm wide is filled to a height $h=87\pm 3~\textrm{mm}$ with a bidisperse mixture of borosilicate glass spheres ($\rho_p =$ 2.23 g/cm$^3$) with diameters $d_l=8~\textrm{mm}$ and $d_s=4~\textrm{mm}$. The larger particles are placed at the bottom, the surface flattened and the smaller particles placed on top. The sidewalls oscillate whilst remaining parallel, applying a periodic shear $\gamma (t) = \gamma_0 \sin(\omega t)$~\cite{scott_bridgwater1975} as shown in Figure~\ref{fig:setup}. The corresponding shear rate $\dot{\gamma}(t) = \gamma_0 \omega \cos(\omega t)$, frequency $\omega = 2\pi/T$~rad\,$\textrm{s}^{-1}$, period $T=13~\textrm{s}$  and strain amplitude $\gamma_0 = \tan\theta_{max}$. The sidewalls displace to a maximum angle $\theta_{max} = \pm30\degree$, giving a maximum shear rate of $\dot{\gamma_0}=\gamma_0 \omega$ and a maximum grain displacement amplitude $A=h\gamma_0$. The angle is decreased to $\theta_{max} = \pm10\degree$ for the particle trajectory data in order to slow down the segregation and increase the temporal resolution. Non-dimensional time $\hat{t}=t/T$ corresponds to the number of elapsed cycles. We follow a sample using RIMS, with the index-matched liquid a mixture of benzylalcohol and ethanol (viscosity $\mu=3~\textrm{mPa\,s}$) containing a fluorescent dye (rhodamine). The low viscosity of the interstitial liquid means that fluid drag forces are small compared to both gravitational forces and the applied shear (Stokes number\,$>$\,1 \cite{cassar2005}). The mixture is lit with a 532~nm laser sheet perpendicular to the oscillating walls, giving a stack of vertical cross sections. A scan is performed after each full oscillation with the shearbox in the upright position. The images were processed using convolution~\citep{shattuck2009} to give three dimensional particle positions, which are coarse-grained in order to determine a continuous volume fraction~\cite{weinhart2013}.  Some sidewall effects exist, with small particles preferentially located near the stationary vertical walls, but this does not affect the overall segregation. The horizontal particle motion is diffusive, hence the concentration is spatially averaged to give a uniform concentration in the $x$-$y$ plane. We observe no convection rolls~\cite{RoyerChaikin2015}, although for $|\phi_{max}|>45\degree$ geometrical squeezing was seen to cause convection.

\begin{figure}[!b]
\resizebox{0.48\textwidth}{!}{  \includegraphics{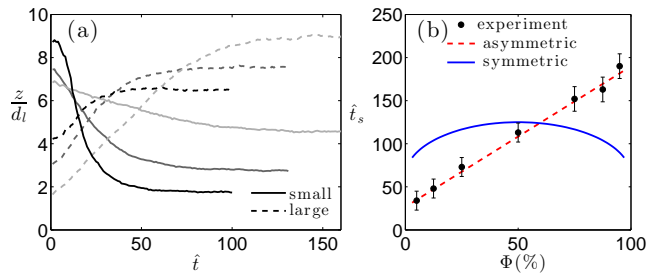}}
\caption{(a) Time evolution of the vertical center of mass position ($\frac{1}{n} \sum_{i=1}^n z_i$) for large and small particles in $\Phi=25\%$ (black), $50\%$ (dark gray) and $75\%$ (light gray) mixtures. $\theta_{max} = \pm30\degree$. (b) Segregation time $\hat{t}_s$ as a function of $\Phi$; solid line is a fit for the symmetric model with $S_r=0.016$, while the dashed line is a fit for asymmetric model with $S_r=0.030$ and $\kappa=0.89$.}
\label{fig:volratio_sketch}
\end{figure}

\emph{Results.---}The typical behavior is shown in Fig.~\ref{fig:setup}: The initial state with large particles on the bottom evolves to a final state with large particles on top, because small particles sink and large particles rise. Interestingly, some large particles remain below when all the others have reached the top. These particles are not stuck but rise at a slower rate than the ones that have reached the top before them. Although this has been inferred before, it has not yet been explained~\cite{golick_daniels2009}.

We define a segregation time $\hat{t}_s$ as the time needed for the vertical centers of mass of the two species to reach a steady state, as shown in Fig.~\ref{fig:volratio_sketch}(a). We record $\hat{t}_s$  for mixtures with varying global volume fraction of small particles $\Phi (\%)= V_s  /(V_{l} + V_{s})$, while keeping the total mixture volume constant. Figure~\ref{fig:volratio_sketch}(b) shows that $\hat{t}_s$ scales linearly with $\Phi$, i.e. with more small particles in the mixture the segregation is slower~\cite{staron2014}. Similar trends were observed over the entire range of angles that can be accessed in our setup. This behavior is usually given a two-part explanation: At low $\Phi$, small particles move slower when there are more small particles~\cite{jha2008}. At high $\Phi$, it takes a longer time for large particles to travel to the top when the layer of small particles above them is thicker~\cite{golick_daniels2009,staron2014}. In both explanations the behavior of the other species is ignored. So how do these explanations combine at an intermediate $\Phi$? A clue is given by~\cite{golick_daniels2009} which reported that for a $\Phi=50\%$ mixture the transition from the state with small particles on top to a mixed state was faster than the subsequent transition from the mixed state to the final segregated state. This points to two separate processes that are likely to be related to the distinct behavior of small and large particles.

\begin{figure}[!b]
  \resizebox{0.48\textwidth}{!}{
    \includegraphics{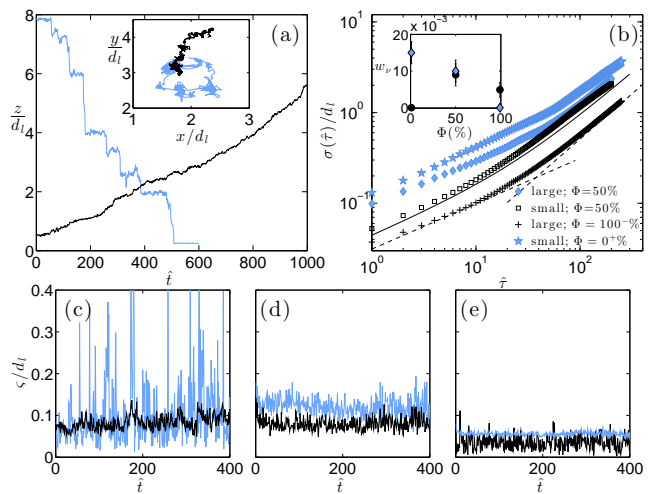}}
\caption{(color online) Individual particle dynamics for small particles (blue, gray) and large particles (black) with $\theta_{max} = \pm10\degree$. (a) Vertical trajectories of a small particle segregating in a $\Phi=0^+\%$ mix; and a large particle segregating in a $\Phi=100^-\%$ mix. Inset: Particle movement in the horizontal plane. (b) RMSD $\sigma(\hat{\tau})$ for different mixtures (see legend), with the solid line a fit of $\sigma_s=\sqrt{D_0\hat{\tau} + w_s^2 \hat{\tau}^2}$ at $\Phi=50\%$ (shifted for clarity). The dotted lines show the slopes 1/4 and 1/2. Inset: $w_\nu(\Phi)$ for large ($\nu=l$) and small particles ($\nu=s$). (c)-(e) $\varsigma $ for single cycles in $\Phi=0^+\%$, $50\%$ and $100^-\%$ mixtures respectively.}
\label{fig:trajectories}
\end{figure}

\emph{Particle dynamics.---}We are thus motivated to study a single small particle segregating in a mixture of large particles and a single large particle segregating in a mixture of small particles, which we refer to as $\Phi=0^+\%$ and $\Phi=100^-\%$ mixtures respectively. The trajectories of the two particles, shown in Fig.~\ref{fig:trajectories}(a), are quite different: (i) the large particle segregates roughly three times slower than the small particle; and (ii) the large particle rises smoothly at an almost constant speed, whereas the small particle shows a stepwise motion with steps of the order of $d_l$. This suggests that the small particle falls through gaps in the large particle matrix under gravity, typically traversing just a single layer. 

\begin{figure}[!b]
\resizebox{0.48\textwidth}{!}{ \includegraphics{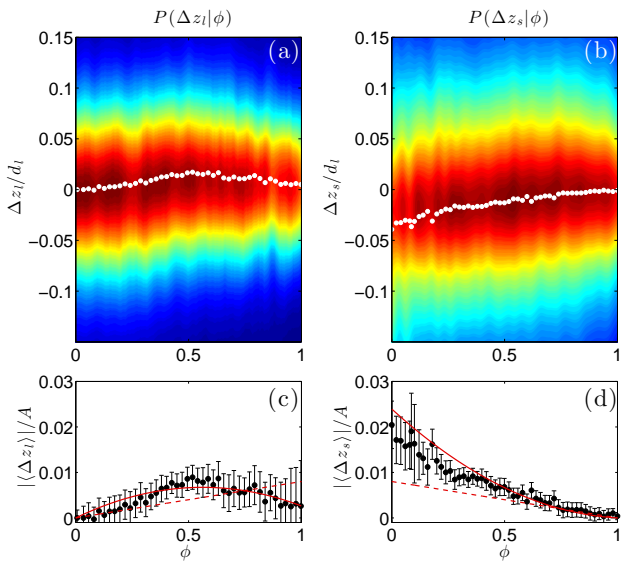}}
\caption{ (color online). (a)-(b) Conditional probabilities $P(\Delta z_l | \phi)$ and $P(\Delta z_s | \phi)$. White curves are $\langle\Delta z_l \rangle$ and $\langle\Delta z_s \rangle$.  (c)-(d) $|\langle\Delta z_l \rangle|/A$ and $|\langle\Delta z_s \rangle|/A$ as a function of $\phi$, with error-bars indicating the standard error of the mean. Dashed and solid lines are plots of Eq.~\eqref{eq:v} for quadratic and cubic flux functions $F(\phi)$ with $S_r=0.008$ and $S_r=0.015$ respectively.  The values of $S_r$ were scaled to account for the lower shear rate $\gamma_0\omega$ at $\theta_{max} = \pm10\degree$.}
\label{fig:statistics}
\end{figure}

In order to more precisely understand the nature of these trajectories, we study the displacement after $\hat{\tau}$ cycles: $\Delta r(\hat{\tau})= r(\hat{t}+\hat{\tau}) - r(\hat{t})$. The root mean square displacement (RMSD) $\sigma(\hat{\tau})=\sqrt{\langle\Delta r^2 (\hat{\tau}) \rangle} $ is plotted in Fig.~\ref{fig:trajectories}(b). The dynamics are diffusive (logarithmic slope~$1/4$) for both particles at short timescales and super diffusive (logarithmic slope~$1/2$) at longer timescales. The crossover length scale between the diffusive and segregation (super-diffusive) regimes for the small particle is approximately $d_l$, which corresponds to the typical segregation step size of the small particle. The crossover length scale for the large particle is lower, roughly $0.2d_l$ ($0.4d_s$), and is likely to be related to the scale of the rearrangements of the surrounding small particles. To confirm this, we measure the RMSD per cycle $\varsigma=\sqrt{\langle\Delta r^2 \rangle} $, as shown in Figs.~\ref{fig:trajectories}(c) and~\ref{fig:trajectories}(e). The typical value of $\varsigma$ for the single large particle lies just below $\varsigma$ for the surrounding small particles [Fig.~\ref{fig:trajectories}(e)]. Although the displacements $\varsigma$ for the single small particle experiences large variations, as a result of falling through layers, the mean value is of the same order as that of the surrounding large particles [Fig.~\ref{fig:trajectories}(c)]. 

The plot of $\sigma(\hat{\tau})$ for a $\Phi=50\%$ mixture in Fig.~\ref{fig:trajectories}(b) shows that the curves lie between those for $\Phi=0^+\%$ and $\Phi=100^-\%$, but  with a comparable amount of segregation. Fitting each of the curves with $\sigma_\nu(\hat{\tau})=\sqrt{D_0\hat{\tau} + w_\nu^2 \hat{\tau}^2}$ with diffusion coefficient $D_0$ allows us to examine the segregation velocities $w_\nu$ for large ($\nu=l$) and small ($\nu=s$) particles at different $\Phi$. The inset of Fig.~\ref{fig:trajectories}(b) shows that $w_s(\Phi)$ decreases with increasing $\Phi$, whereas $w_l(\Phi)$ increases to a maximum at $\Phi=50\%$ and then decreases, although not to zero, at $\Phi=100^-\%$. To understand the peak in $w_l(\Phi)$, we plot $\varsigma$ for $\Phi=50\%$ in Fig.~\ref{fig:trajectories}(d). The values of $\varsigma$ for both small and large particles increase with respect to the $\Phi=100^-\%$ mix, however, the large particle movement is still less compared to the small particles. 

At this point we can hypothesize an explanation for the trend in Fig.~\ref{fig:volratio_sketch}(b): the individual dynamics of small and large grains have a different significance on the overall segregation dynamics at different $\Phi$. At high $\Phi$, the significant dynamics are of the `slow' large particle, which are governed by the scale of rearrangements of the surrounding small particles. At low $\Phi$, it is the `fast' small particle which is significant as it can make big segregation steps between large particle layers.  At an intermediate $\Phi$ both processes combine; small particles slow down, because layering disappears, while large particles speed up, because the scale of rearrangements increases.

\emph{Displacement statistics.---}To study this behavior at the particle scale for each species ($\nu=l,s$), we measure the conditional probabilities $P(\Delta z_\nu| \phi)$ of the vertical displacement $\Delta z_\nu$ given that the local small particle volume fraction is $\phi$. Note that shear-gradients~\cite{fan_hill2011} do not play a role, because of the linear shear profile that is applied. Here, $\phi=0$ corresponds to regions of only large particles and $\phi=1$ to only small particles. The results in Figs.~\ref{fig:statistics}(a) and \ref{fig:statistics}(b)  demonstrate that large particles are less likely to segregate at high $\phi$ compared to  small particles segregating at low $\phi$. In the following, we will refer to this as ``asymmetry''. Similar to the data for $w_l(\Phi)$ in the inset of Fig.~\ref{fig:trajectories}(b), we observe that the large particles have their greatest displacement at an intermediate $\phi$.

\begin{figure}[t]
\begin{center}
\resizebox{0.48\textwidth}{!}{\includegraphics{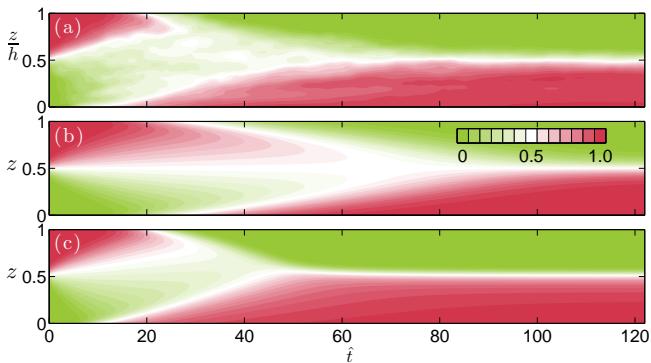}}
\caption{(color online). (a) Temporal development of $\phi(z,\hat{t})$ versus normalized flow height $z/h$ for a $\Phi=50\%$ mixture with $\theta_{max} = \pm30\degree$. (b)-(c) Theoretical predictions from Eq. \eqref{eq:segregationequation}. (b) Prediction using the symmetric flux function~\eqref{eq:quadraticflux}, with $S_r=0.016$ and $S_r/D_r=20.9$~\cite{fitting}. (c) Prediction using the asymmetric flux function~\eqref{eq:cubicflux}, with $S_r=0.030$, $S_r/D_r=29.6$ and $\kappa=0.89$.}
\label{fig:result}
\end{center}
\end{figure}

The effect of asymmetry at a mesoscale can be seen in the temporal development of $\phi(z,\hat{t})$ for a $\Phi=50\%$ mixture in Fig.~\ref{fig:result}(a). Two important features exist: (i) small particles reach the bottom of the flow faster compared to large particles reaching the top; (ii) large particles appear to rise predominantly together (indicated by the band of low $\phi$). The first feature is easily explained by asymmetry: small particles beginning the experiment near the interface between the two species quickly travel to the bottom through the large particle matrix, in accordance with $P(\Delta z_s | \phi)$. The second feature is possibly linked to the large particles having a maximum segregation speed at an intermediate concentration.  

\emph{Theoretical comparison.---}Current approaches to modeling size segregation use an advection-diffusion equation for $\phi$~\cite{GrayGajjarKokelaar2015}:
\begin{equation}
 \dfrac{\partial \phi}{\partial t} + \text{div}(\phi \mathbf{u}) - \dfrac{\partial}{\partial z}\bigl(qF(\phi)\bigr)=\dfrac{\partial}{\partial z}\bigl(D\dfrac{\partial \phi}{\partial z}\bigr), \label{eq:segregation-advectiondiffusionequation}
\end{equation}
where $\mathbf{u}$ is the bulk velocity field, $q=q(\omega\gamma_0,g)$ is the mean segregation speed, $g$ is gravity, $D$ is the diffusivity and $F(\phi)$ is the flux function, which determines the dependence of the segregation flux on $\phi$. The simplest flux function has a quadratic form
\begin{equation}
 F(\phi)=\phi(1-\phi). \label{eq:quadraticflux}
\end{equation}
This is employed in a number of models~\cite{gray_thornton2005, may2010, fan2014} and is symmetric about $\phi=0.5$, dictating that small and large particles behave identically, but in opposite directions. Recently, asymmetric flux functions were introduced by \cite{gajjar_gray2014} with the simplest being a cubic form
\begin{equation}
 F(\phi)=A_\kappa\phi(1-\phi)(1-\kappa\phi), \label{eq:cubicflux}
\end{equation}
where asymmetry parameter $0\leqslant\kappa<1$, and normalization constant $A_\kappa$ gives the same amplitude as the symmetric flux function. 
The applied shear gives a velocity profile $\mathbf{u}=(u(z,t),0,0)$. In combination with the lateral spatial uniformity of $\phi$, this means that the transport term in Eq.~\eqref{eq:segregation-advectiondiffusionequation} is zero. Equation~\eqref{eq:segregation-advectiondiffusionequation} reduces to:
\begin{equation}
  \dfrac{\partial \phi}{\partial t}  - \dfrac{\partial}{\partial \hat{z}}\bigl(S_rF(\phi)\bigr)=\dfrac{\partial}{\partial \hat{z}}\bigl(D_r\dfrac{\partial \phi}{\partial \hat{z}}\bigr), \label{eq:segregationequation}
\end{equation}
where $\hat{z} = z / A$, and $S_r=qT/A$, $D_r=DT/A^2$ are non-dimensional segregation and diffusive-remixing coefficients respectively.
The symmetric and asymmetric models were least squares fitted to the data in Fig.~\ref{fig:volratio_sketch}(b) to obtain $S_r=0.016$ for the symmetric model and $\kappa=0.89$, $S_r=0.030$ for the asymmetric model. Integrating Eq.~\eqref{eq:segregationequation} gives the $\phi$ evolution in Figs.~\ref{fig:result}(b) and~\ref{fig:result}(c). Qualitatively, Fig.~\ref{fig:result}(c) reproduces the experimental result on some critical points: (i) the difference in time between the arrival of small particles at the bottom and large particles at the top of the flow; (ii) the collective rising of large particles; and (iii) a lower $\phi$ in the bottom half of the flow near the end of the experiment, indicating that some large particles are still inside the small particle matrix, segregating very slowly. These features are not found in the symmetric result in Fig.~\ref{fig:result}(b).
The theoretical displacements per cycle are given by
\begin{equation}\label{eq:v}
 \bigl|\Delta \hat{z}_l\bigr| = S_r\dfrac{F(\phi)}{1-\phi},\qquad \bigl|\Delta \hat{z}_s\bigr| = S_r\dfrac{F(\phi)}{\phi},
\end{equation}
and are shown alongside the experimental data in Figs.~\ref{fig:statistics}(c) and~\ref{fig:statistics}(d). The trend is clearly better predicted by the asymmetric flux, which is able to reproduce both the peak in $|\langle \Delta z_{l} \rangle |$ around $\phi=0.5$ and the nonlinear decrease of $|\langle \Delta z_{s} \rangle |$. We attribute the discrepancy of $|\langle \Delta z_{s} \rangle |$ at low $\phi$ to tracking errors, when small particles move more than their radius and their displacement is not recorded, thereby lowering the measured value. 

\emph{Discussion.---}We analyze particle motion in a segregating bidisperse mixture under oscillatory shear and discover an underlying asymmetry in the behavior of large and small particles. The small particle motion is step-like, falling down through the large particle matrix typically one layer at a time. On the other hand, the large particle motion is smoother but slower, and linked to the scale of rearrangements of the surrounding small particles. The asymmetric motion of the large and small particles combine to give a characteristic dependence of the particle segregation speeds on the local volume fraction. Large particles segregate slower in the presence of many small particles, while small particles segregate faster in the presence of many large particles. We also observe that large particles move quickest when close to other large particles at intermediate concentrations, a process reminiscent of a collective motion~\cite{vicsek2012}. The underlying asymmetry also manifests at meso and bulk scales. In the development of $\phi(z,\hat{t})$, small particles reach the bottom of the flow faster than large particles reach the top. The segregation time increases linearly when a mixture contains a larger fraction of small particles. Although there is no direct evidence that the observed asymmetry persists for continuous shear, \citet{staron2014} report that the segregation time under steady shear also increases linearly with the total concentration of small particles. These insights give a new understanding of segregation in sheared systems, with the dynamic behavior of two species being  inherently different.  

Models for segregation have typically considered the motion of the large and small grains to be identical. However, an experimentally determined cubic flux \cite{gajjar_gray2014} brings asymmetric behavior for the two species and gives good agreement on both particle and bulk scales.  This draws parallels with the use of asymmetric flux functions to model asymmetry in sedimentation~\citep{batchelor1972}, traffic flows~\cite{LighthillWhitham1955b,TRB2011} and diffusion across membranes~\cite{Packard2004}. For example, in the sedimentation of suspensions, particles settle faster when traveling together, but the settling velocity goes to zero more rapidly than a linear decrease at high concentrations~\citep{batchelor1972}. Similarly, the velocity of cars in traffic approaches zero non-linearly at high car densities~\cite{TRB2011}. The commonality between these processes is their discrete nature, but interestingly, size segregation is the only process that consists of two discrete species.

Further work is needed to analyze particle scale motion for continuously sheared flows, e.g. down chutes and within rotating drums, to determine whether asymmetry persists and what form the segregation flux takes. The distinct segregation dynamics of the two species also leads to questions as to a possible relation with other dynamic processes such as dynamic heterogeneity~\cite{candelier2009}.

\emph{Acknowledgements.---}The authors would like to acknowledge enlightening discussions with Marco Ramaioli, Karen Daniels, Anthony Thornton and Matthias Schr\"oter, as well as technical and analytical assistance from Justine Caillet. Financial support for this work came from the Swiss SNF grant 200021-149441 and NERC grants NER/A/S/2003/00439 and NE/E003206/1, as well as EPSRC grants EP/I019189/1 and EP/K00428X/1.


\begin{thebibliography}{48}
\bibitem[{fit()}]{fitting}
\bibinfo{note}{The ratio $S_r/D_r$ is found by least squares matching the
  steady end state.}
\end{thebibliography}
\end{document}